\documentclass[a4paper]{article}

\usepackage[top=1.5in, bottom=1.5in, left=1in, right=1in]{geometry}

\usepackage{amsmath}
\usepackage{multirow}
\usepackage{url}
\usepackage{booktabs}
\usepackage[tight]{units}
\usepackage{graphicx}
\usepackage{epstopdf}
\usepackage{hyperref}
\usepackage{authblk}

\DeclareMathOperator{\AR}{AR}
\DeclareMathOperator{\AL}{AL}

\begin{document}

\title{TreQ-CG: Clustering Accelerates High-Throughput Sequencing Read Mapping}

\author[Mahmud and Schliep]{Md Pavel Mahmud\,$^{1}$\footnote{to whom correspondence should be addressed}\, and Alexander Schliep\,$^{1,2\,*}$ \\
$^1$Department of Computer Science, Rutgers University, NJ, USA.\\
$^2$BioMaPS Institute for Quantitative Biology, Rutgers University, NJ, USA..}

\maketitle

\section{Abstract}

  As high-throughput sequencers become standard
  equipment outside of sequencing centers, there is an increasing need
  for efficient methods for pre-processing
  and
  primary analysis.
  While a vast literature proposes methods for
  HTS data analysis, we argue that significant improvements can
  still be gained by
  exploiting expensive pre-processing steps which can be amortized with
  savings from later stages.
  We propose a method
  to accelerate and improve read mapping
  based on an initial
  clustering of possibly billions of high-throughput sequencing reads,
  yielding clusters of high stringency and a high degree of overlap.
  This clustering improves on the state-of-the-art in running time
  for small datasets and, for the first time, makes
  clustering high-coverage human libraries feasible.
  Given the
  efficiently
  computed clusters, only one representative read
  from each cluster needs to be mapped using a traditional readmapper
  such as BWA, instead of individually mapping all reads.
        On human reads, all processing
  steps, including clustering and mapping, only require 11\%--59\% of the
  time for individually mapping all reads,
  achieving speed-ups for all readmappers,
  while minimally affecting mapping quality.
  This accelerates a
  highly sensitive readmapper such as Stampy to be competitive with
  a fast readmapper such as BWA on unclustered reads.

\paragraph{Availability:}
        TreQ-CG is freely available for download at
  \url{http://bioinformatics.rutgers.edu/Software/TreqCG}.



\section{Introduction}

The amount of data produced by current high-throughput DNA
sequencing machines such as Illumina HiSeq 2500, which can generate as much as 100Gb a day, demands enormous computational power for primary analysis tasks
such as read mapping.
Although a large body of literature is concerned with read mapping \cite{Li01082009, Li15072009, Langmead:2009fv, Rumble:2009jw, Hoffmann:2009hr, Hach:2010cd, Lunter2011, Hamada:2011bg, Weese:2012by, Mahmud:2012jn, Ahmadi01032012},
most approaches map one read at a time. The order of mapping is arbitrary regardless of similarities between reads which might hint towards the same mapping location. One notable exception is Masai \cite{Siragusa2013}, which jointly maps read prefixes using a radix tree, but necessarily has to rely on exact prefix matches.
In high coverage libraries, reads originating from the same genomic locus
share many bases, which, if processed together,
can significantly reduce the computational effort.
However, identifying partial overlap in the presence of sequencing errors,
which are artifacts of library construction and other experimental errors,
is itself a computationally expensive task.

From a computational point of view, read mappers  can  roughly be divided in two categories:
Programs such as BWA \cite{Li15072009} are very fast at mapping reads with low edit distance to the reference, but become exponentially slower with increasing edit distance. They often  have an internal cut-off to keep running times in check. More sensitive programs, such as Stampy \cite{Lunter2011}, can handle larger edit distance in the presence of a
high degree of variation, as found in cancer datasets \cite{Shah:2012cea}, at the price of being slower in general, typically by one order of magnitude.
Interestingly, those two categories roughly correspond to the two sources of edit operations observed in HTS reads: sequencing errors and intergenomic variation. BWA is suitable for mapping reads from loci in the sample genome which are almost identical to the corresponding locus in the reference, and the edit distance is mostly due to a few sequencing errors. However, for the case of larger variation between sample and reference, read mappers such as Stampy are tailored towards large edit distance. In that case, mapping reads individually means the algorithm has to handle the difference  between loci in sample and reference as many times as there are reads covering it, which is wasteful. As the difference between the overlapping segments of reads from the same
genomic locus is solely due to sequencing errors, their mutual edit distance
is usually {\em less} than their edit distance to the reference
genome in the presence of a high degree of intergenomic variations. Consequently, identifying similar reads that most likely come from the same sample locus and mapping them jointly is faster than individual mapping. Unfortunately, Stampy and other sensitive readmappers do not utilize that observation.

\begin{figure}
\centering
\includegraphics[scale=0.75]{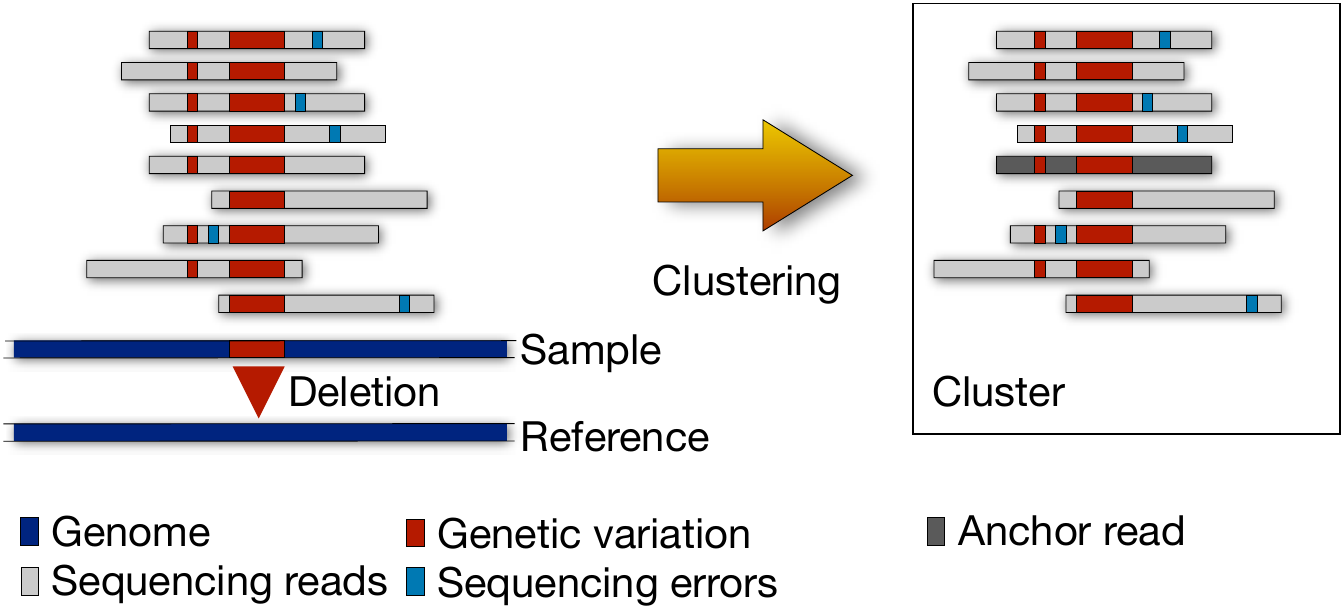}
\caption{\label{fig:method}
{\bf Schematic view of clustering.}
Genetic variations do not contribute to the edit distance between reads from the same genomic
location. Thus, the only source of difference in an overlap is sequencing error. Since the sequencing
error rate is small for Illumina-like datasets, in most cases, the clustering process can overcome
the differences in the overlap between reads and group them together in a cluster. Moreover, to
reduce the probability of assigning reads from different genomic locations to the same cluster,
the overlap length has to be sufficiently large.}
\vspace{-4mm}
\end{figure}

The first step in exploiting the redundancy among reads originating
from the same genomic locus is to identify them through clustering.
To compute clusters, we use an incremental greedy approach
ensuring stringent criteria for cluster membership in terms of overlap length
and similarity to the anchor read representing the cluster (see Figure~\ref{fig:method}).
The clustering we use is unusual in two aspects: By design our number of clusters is very large (5-10\% of the number of reads) and different clusters can contain reads from the same genomic locus. Both conditions are necessary to minimize mis-mappings due to clustering.
Prior works on clustering HTS reads were concerned with detection
of clusters in which members either completely or almost fully
overlap with a high degree of similarity between them \cite{Bao:2011ia,Shimizu:2011it,Edgar01102010}.
Relaxing that criterion makes clustering computationally feasible,
and it is not detrimental to further analysis as the clusters are not per
se focus of the investigation.
Assuming that one cluster contains reads sequenced
from the same genomic locus, we only map one anchor
read per cluster and align other cluster members using a local alignment
algorithm. Mapping clusters instead of individual reads speeds up
computation irrespective of the readmapper used.

Since clustering large datasets
is a fundamental pre-processing step in data analysis, it has been  used extensively for many HTS-related
problems.
For example, in read compression, some tools exploit redundancy among reads to
achieve higher compression rate without using a reference genome; e.g.,
Coil \cite{18489794} and ReCoil \cite{21988957} use an approach similar to
us for grouping similar reads together before compression.
Similar ideas have also been applied to multiple sequence alignment-based
error correction tools such as Coral \cite{citeulike:9110954} and ECHO \cite{citeulike:9144429}.
Redundancy removal is another application based on clustering, e.g., SEED \cite{Bao:2011ia}
shows that removing redundant reads can improve running time, reduce memory
requirement and improve de-novo assembly quality. Similarly, for RAD-Seq data,
Rainbow \cite{ChongRW12} shows improved de-novo assembly after clustering.
Although many diverse HTS tools exploit sequence redundancy through clustering, 
there are very few tools that use clustering as a pre-processing step before
read mapping. One such tool, FreeClu \cite{Qu01072009}, creates an interesting parent-child
tree structure based on read frequency and Hamming
distance between reads. They report increased read mapping rate by utilizing
the relationship among reads represented as a tree. Unfortunately, their algorithm is not
scalable to large datasets, and also restricted to full overlap between reads allowing a Hamming distance of at most one.
Another tool, Oculus \cite{23148484}, combines redundancy check and read mapping together in one package.
They process one read at a time in a streaming setting,  align a  previously unseen read using Bowtie, and store results in a hash table, thereby skipping expensive alignment step for redundant reads.
Like FreClu, Oculus needs full overlap 
and, unlike FreClu, requires complete match between reads to declare them redundant.
This approach can achieve moderate speed-up for very high coverage RNA-Seq data,
but is unlikely to be useful for whole genome sequencing data in presence of sequencing errors. In contrast, we process
up to a billion reads in one batch, and allow reads to partially overlap as well as contain
mismatches in the overlap (similar in spirit to MSA-based error correction approaches),
which leads to significant speed-up in read mapping.


\section{Methods}
By $R = \{r_1, \ldots, r_N\}$ we denote the read library.
We assume all reads to have a fixed length $L$.
We also assume $R$ to be produced by an Illumina-like sequencing experiment where
substitution errors are dominant compared to insertion and deletion (indel)
errors. In the following, we discuss how we efficiently cluster $R$
and use these clusters to accelerate read mapping.

\subsection{Clustering billions of reads for mapping}
\label{clustering}
Our two main design goals for the clustering are computational efficiency
and stringency with respect to
assigning reads from the same genomic locus
to one cluster. Stringency and the size of the data imply large number of
clusters, which precludes the use of iterative clustering methods. Note
that the clusters themselves are not the focus of investigation, but rather a
computational aid. Consequently, we allow reads from the same locus to be
assigned to distinct clusters. We propose the following greedy approach based
on shared $k$-mers for clustering reads.

\subsubsection{Clustering single-end reads}
Given two reads $r_i$ and $r_j$ with a prefix-suffix overlap of length $l$,
where $l'$ number of bases match, we say that they have an overlap of
length $l$ with similarity $s = \frac{l'}{l}$.
If $l \ge \alpha L$ and $s \ge \beta$ ($\alpha$ and $\beta$ are constants),
we assume $r_i$ and $r_j$ are \emph{sufficiently similar} to be member of
the same cluster. Clusters are represented by their anchor reads (in
contrast to centroids in $k$-means clusters' anchors remain fixed) and reads,
which are sufficiently similar to the anchors, are greedily assigned to the clusters (Figure~\ref{fig:method}).
If a read of sufficient quality---i.e., not a
{\em bad}
read as defined in the following---fails to
find a sufficiently similar anchor it becomes an anchor read itself and
thus forms a new cluster.

Specifically, we create a set of anchor reads $\AR$ and an array
of $4^{k}$ lists $\AL=\{\AL_0, \ldots, A_m, \ldots, \AL_{4^k-1}\}$,
where $A_m$ stores pointers to the anchor reads that contain the $k$-mer $m$
(note that there is a one-to-one relation between $k$-mers and $2k$ bit numbers),
and perform the following steps for each read $r_i$ in the library.

\begin{enumerate}
  \setlength{\itemsep}{1pt}
  \setlength{\parskip}{0pt}
  \setlength{\parsep}{0pt}

\item A base $r_{i,j}$ is defined as a bad base if it is ambiguous
(represented by 'N') or it has Phred score below the threshold
$f$. If there are too many bad bases uniformly distributed in $r_i$
we call $r_i$ a bad read and discard it from clustering.
In particular, we discard $r_i$ if

\begin{itemize}
\item $r_{i,s}$ and $r_{i,e}$ are the first and last non-bad bases
respectively, and $e-s+1 < 2k$, or


\item the maximum length of a contiguous sub-sequence containing no bad base is less than $\frac{k}{2}$.

\end{itemize}

\item For each $k$-mer in $r_i$, we

  \begin{enumerate}
  \setlength{\itemsep}{1pt}
  \setlength{\parskip}{0pt}
  \setlength{\parsep}{0pt}
  \item compute the corresponding $2k$-bit integer
  representation $m$ and get the list of anchor reads that share
  the same $k$-mer from $\AL_m$.
  \item For each anchor read $r_j$ in $\AL_m$,
  whose content is already stored in $\AR$,
  if $r_i$ and $r_j$ has a prefix-suffix overlap of at least $l \ge \alpha L$ bases
  respecting the position
    of the shared $k$-mer, we compute similarity between $r_i$ and $r_j$
    in the overlapped region.
  \item If at least $\beta l$ number of bases match
  in the overlapped region,
  we declare $r_i$ as a member of the cluster formed around the
    anchor read $r_j$ and halt further computation for $r_i$.
  \item Otherwise, we proceed with the next $k$-mer.
  \end{enumerate}

\item If $r_i$ fails to be assigned to any cluster, it
qualifies to be an anchor read and forms it's own cluster. We store $r_i$
in the set of anchor reads $\AR$ and
for every $k$-mer $m$ present in $r_i$'s
$\alpha L$-length prefix and suffix we insert a record in $\AL_{m}$.
\end{enumerate}

\subsubsection{Clustering paired-end reads}
Although clustering a paired-end read library in single-end mode is
acceptable for many applications, for paired-end read mapping
it is necessary to take pairing information
into account.
Unlike single-end clustering, where we consider one read at a time,
in paired-end clustering, we process a pair of reads simultaneously
and do not allow one end of a pair to be anchor read and the other
end to be non-anchor read.
This constraint lets us map anchor reads in paired-end
mode using a traditional readmapper.
If the constraint is violated for a pair $r_{i_1}$ and $r_{i_2}$,
where, lets say, $r_{i_1}$ is an anchor read but $r_{i_2}$ is not,
we resolve it in the following order.
\begin{enumerate}

\item If $r_{i_2}$ is a bad read, we force it to be an anchor read and
form a new cluster.
\item If $r_{i_2}$ is a member of the cluster formed by anchor read
$r_{j_2}$, whose other end is another anchor read $r_{j_1}$,
we try to find a prefix-suffix overlap of length
$l \ge \frac{\alpha L}{2}$ between $r_{i_1}$ and $r_{j_1}$ where
$\beta l$ bases match (since $r_{i_2}$ already provides some evidence
for this choice we relax the overlap length restriction from
$\le \alpha L$ to $\le \frac{\alpha L}{2}$).
If such an overlap is found we force
$r_{i_1}$ to be a member of $r_{j_1}$.
\item Otherwise, $r_{i_2}$ is forced to be an anchor read and form a
new cluster.
\end{enumerate}

\subsubsection{Optimal cluster assignment}
The above procedure partitions the read library into three sets; bad reads, anchor
reads and member reads.
Since not all anchor reads were known before the completion of the first phase,
we defer final cluster assignment to the second phase of the algorithm.
In particular, for each member read $r_i$ with an overlap of length $l$
with it's current cluster's anchor read, we try to change the
assigned cluster by finding another anchor read with an overlap $l' > l$.
We perform this by running only step 2 from the single-end algorithm
and modifying step 2.c to find the best possible cluster assignment.

\subsubsection{Sub-optimal cluster assignment for repeat resolution}
\begin{figure}
\centering
\includegraphics[scale=0.5]{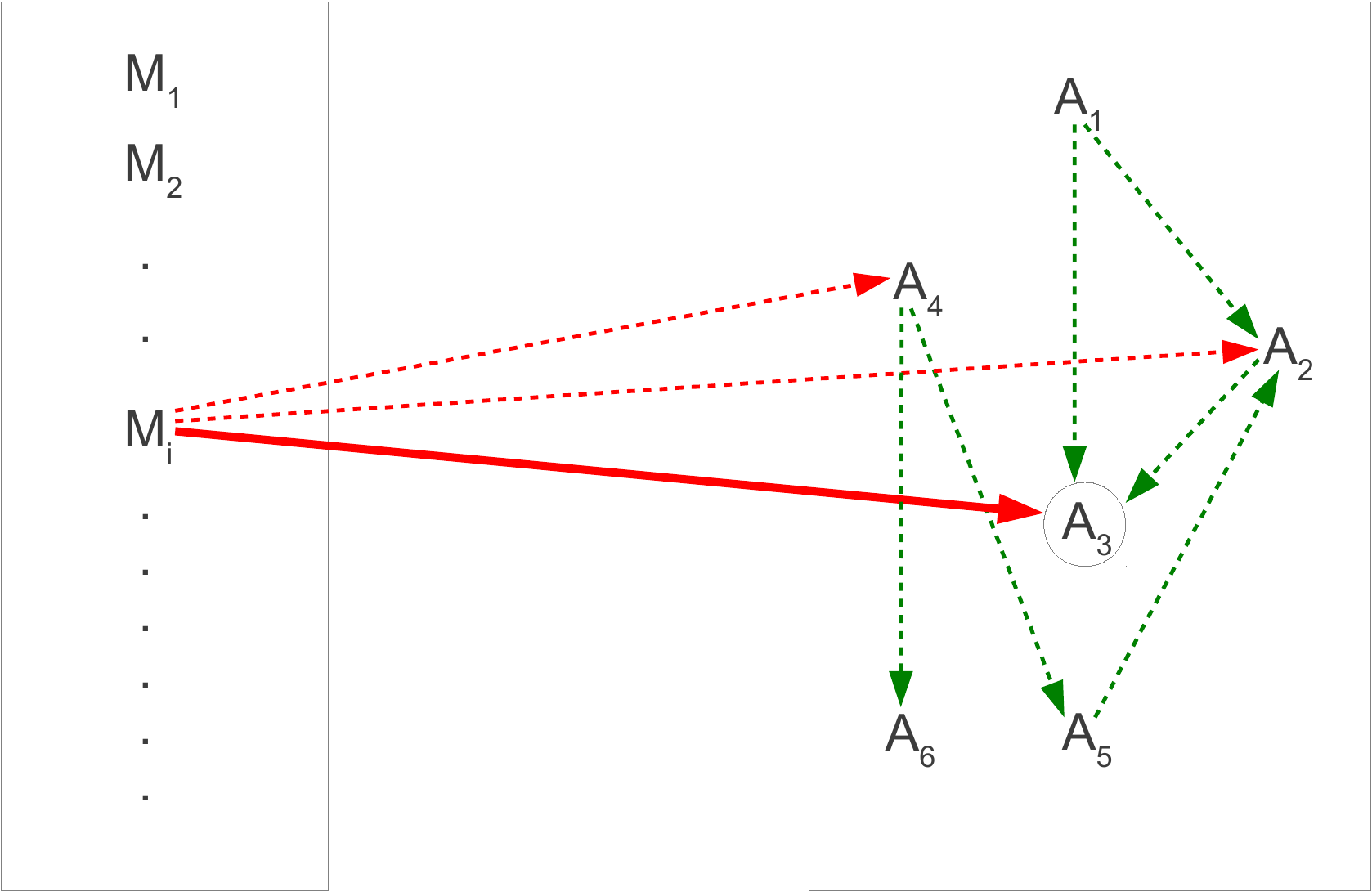}
\caption{\label{fig:repeat}
{\bf Sub-optimal cluster choices.}
Here $M_i$ is a member read of the cluster formed by anchor read $A_3$
  (relationship shown with a solid red arrow).
  In addition to $A_3$, $M_i$ also has at least $\alpha L$ length overlap with
  $A_2$ and $A_4$, where at least ($\beta'\times\mbox{overlap length}$) number of bases match
  (shown with dotted red arrow).
  Similarly, some of the anchor reads also have significant overlap
  among them (shown with dotted green arrow). These sub-optimal choices
  are considered in the read mapping step to increase the probability of
  accurately mapping non-unique, repetitive reads.
}
\vspace{-4mm}
\end{figure}
Another crucial aspect to take into consideration is the complexity of
the underlying genome from which sequencing reads originate. Because
of the non-uniformity and repetitiveness present in the genome,
sub-optimal cluster assignments are sometimes preferred. Towards this goal,
in the second phase, along with the cluster with highest overlap, we also
store at most $S_M$ number of sub-optimal (allowing $\beta' < \beta$)
cluster assignments for each member read (red dotted arrows in Figure \ref{fig:repeat}).
We take this idea further and, for an anchor read, store at most
$S_A$ number of similar anchor reads by modifying the first phase of the
algorithm (green dotted arrows in Figure \ref{fig:repeat}).
Storing these two kinds of sub-optimal choices
can be very useful in resolving ambiguity and repeats in read mapping.

\subsubsection{Practical Considerations}
Even our greedy two stage
clustering strategy can become infeasible already for one billion
reads. We have carefully taken some decisions and made some parameter choices
to keep the clustering process feasible. In particular, smaller values of
$k$ increase sensitivity but require more
computational resources due to more frequent $k$-mer hits.
Since our cluster definition is comparatively
relaxed, a large $k$, in this case 15,
is suitable for read libraries with few sequencing errors such as
Illumina's.
Additionally, we have made the following choices:
One, we ignore indels introduced by sequencing and perform a Hamming distance computation on the overlap between two reads
only if they share at least two $k$-mers,
which decreases the probability of spurious hits due to the choice of $k$
and the presence of sequencing errors;
two, since overlapping $k$-mers in a read are not independent and,
hence, considering all $k$-mers increases computational demand without significant
improvement in optimal cluster assignment,
in the second phase, we abandon the search for best cluster assignment
after considering $\frac{L-k+1}{4}$ equally spaced $k$-mers from the read;
three, since frequent $k$-mers are less informative, the size of the list
$\AL_m$, storing the anchor reads containing $m$, is restricted to 256;
four, we restrict the number of sub-optimal choices $S_M$ to 16 and
$S_A$ to 256
and store them in external memory;
finally, we set $\beta=0.95$, $\beta'=0.8$ and
$\alpha=\max\{0.5, \frac{31}{L}\}$,
which restricts the allowed read length to be at least 31 and guarantees
that the clustering algorithm can overcome at least one error
$(2\cdot15+1=31)$ in the overlap. See supplement
for a detailed discussion on choosing parameters $\alpha$ and $\beta$.

\subsubsection{Running time and memory usage}
For each of the $N$ reads of length $L$ in the library,
the clustering process computes Hamming distance between the read and
potential anchor reads found using shared $k$-mers.
Consequently, the running time complexity of the algorithm is $O(NL^2)$.
Due to the practical choices from the previous section, in practice,
the total running time is not quadratic in $L$, and the average
running time is approximately $O(NL)$.
For space complexity, lets assume the clustering algorithm finds
$\tau N$ clusters, where $0 \le \tau \le 1$.
In that case, the set of 2-bit encoded anchor reads $\AR$ needs
$O(\tau N L)$ bytes, the array of lists $\AL$ needs
$O(\tau N L + N)$ bytes and cluster membership information takes $O(N)$
bytes. Thus, the total
amount of space required by these three objects is $O(\tau N L)$.
Including other overhead, specially for
multi-threading support, a $100$-bp $1.5$ billion human read
dataset, where $\tau=0.08$ ($\tau=0.11$) for single-end
(paired-end), requires $76$GB ($91$GB) memory.
If the read sequences were error free, we would need significantly less
amount of memory since $\tau$ would be $O\big(\frac{G}{L}\big)$.
But, unfortunately, HTS libraries contain frequent sequencing errors,
and hence, this large space requirement is unavoidable for de-novo clustering.

\subsection{Mapping clusters}

\begin{figure}
\centering
\includegraphics[scale=0.8]{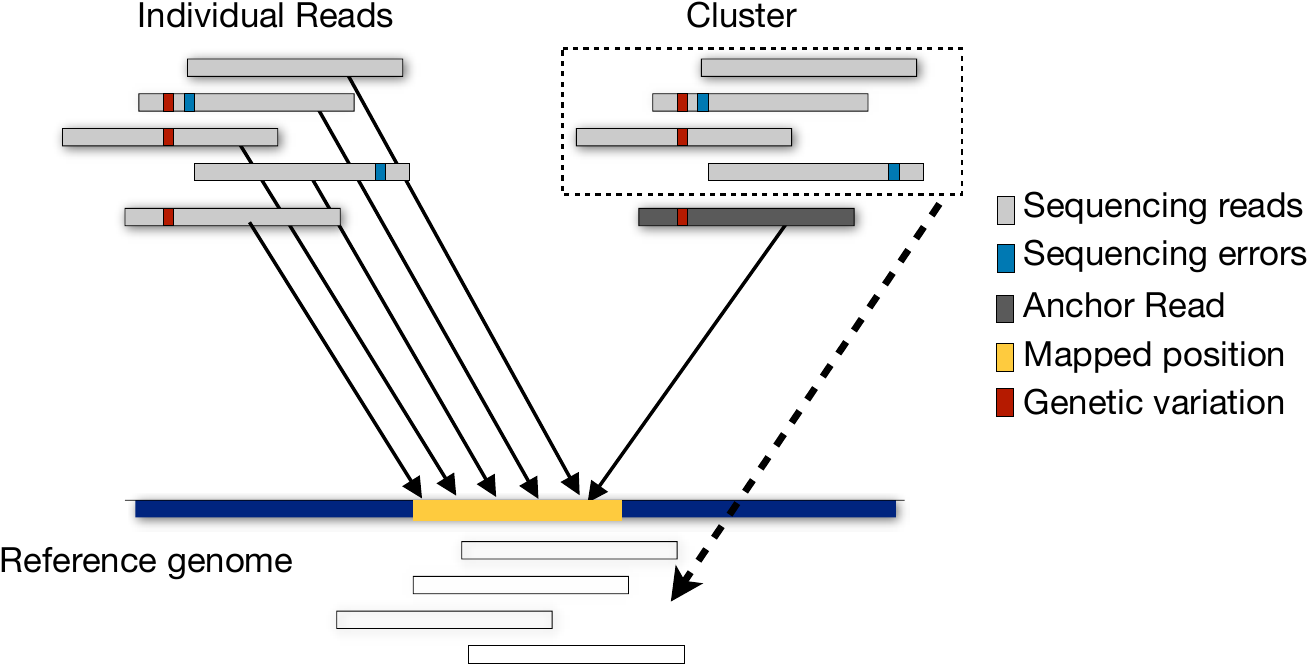}
\caption{\label{fig:mapping-method}
{\bf Schematic view of clustered read mapping.}
Read mappers map all reads in input individually to the reference genome
  (ignoring savings achieved using index structures). In contrast,
  in clustered read mapping only the anchor read is mapped to the
  reference genome. The mapping positions of individual cluster members
  are recovered w.r.t. to the position of the anchor read.
}
\vspace{-4mm}
\end{figure}

In resequencing experiments the first step in the analysis typically
is mapping reads to a reference genome. The choice of a tool
depends on the biological question under investigation, the expected
genetic variation between sample and reference genome and the
computational resources available. The tools roughly fall into two
categories: fast read-mappers which trade losses in mapping
reads
for speed such as
SOAP2 \cite{Li01082009},
Bowtie 2 \cite{Langmead2012} or
BWA \cite{Li15072009}
or highly-sensitive, but much slower tools such as
mrFAST \cite{Alkan_mrFAST},
RazerS \cite{Weese01092009},
LAST \cite{Frith01042010},
Novoalign (\url{http://www.novocraft.com}) or
Stampy \cite{Lunter2011}.
We will show
how we accelerate read mapping based on the clustering irrespective of
the readmapper used.

Our clustering algorithm works under the assumption that
members of a cluster originate from the same genomic locus in the
sample genome. We exploit this assumption by mapping the anchors in
single-end or paired-end mode to a reference genome and subsequently
use the
information about the mapped position of the anchor read to find good
mappings for member reads. As a result the running time complexity of read
mapping reduces to the proportion of clusters found; in practice
we need more time
as the cost of mapping a single read varies
considerably.
Although we can find more than one location for non-unique, repetitive reads
using multiple hits for the anchor reads and utilizing sub-optimal cluster choices,
in this study, for a suitable comparison between clustered vs. individual
read mapping, we only report single best hits for reads.
Specifically, given the read library $R$ we perform the
following steps to compute the read alignments assuming fixed costs
$3$, $-3$, $-40$, and $-3$
for nucleotide match, mismatch, gap open, and gap extension.
\begin{enumerate}
\setlength{\itemsep}{1pt}
  \setlength{\parskip}{0pt}
  \setlength{\parsep}{0pt}
\item Identify the anchor reads and map them in single best mode using
a readmapper. Report the alignments in a SAM file $S$.
\item Extract mapping information of the anchor reads from $S$.
\item For each read $r$ in $R$,
  \begin{enumerate}
  \setlength{\itemsep}{1pt}
  \setlength{\parskip}{0pt}
  \setlength{\parsep}{0pt}
  \item If $r$ is an anchor read, do nothing.
  \item If $r$ is a bad read, report the read as unmapped in $S$.
  \item If $r$ is a member read,
	\begin{itemize}
	\item we use the overlap information between anchor read and member
	read to identify the genomic position from where
	the member read may have originated, and compute Hamming distance.
	\item If the total cost of mismatches from previous step is greater
	than a single gap opening
	cost (for the tested datasets it happens for approximately 30\% of the reads)
	we run the Smith-Waterman
    algorithm \cite{Smith1981195} to find the best local alignment
    between the read and the mapped position of the anchor in the genome
    ($\pm L$ bases). We use a modified version of a SSE based library~\cite{Beneke:1997hv} for Smith-Waterman computation.
    \item We repeat the above two steps for the sub-optimal cluster
    choices (Figure~\ref{fig:repeat}). In particular, we increase the search
    space by following a path, through the dotted red and green links, of length at most two,
    from the member read to the anchor reads.
	\item If the best alignment has score less than {\tt min\_score}
	(default $\frac{L}{3}$) we report the read as
    unmapped. Otherwise, we append the alignment information at the
    end of file $S$.
    \item For paired-end reads, if pairs are concordant we do nothing.
    For discordant pairs, we use one end's mapping information to find
    a better alignment for the other end before reporting them in $S$.
    \end{itemize}
  \end{enumerate}
\end{enumerate}

Since readmappers often use different kinds of alignments --- choice varies
among local, global, and semi-global alignment,
score values, heuristics, and strategies to improve
paired-end mapping, it is difficult to
select an universal set of criteria which is good for all. The
differences in choices and heuristics used give them distinguishing
characteristics which is often desirable by downstream analysis tools.
We believe that the choices we have made here are not
necessarily the best for every readmapper,
and it makes sense to have different clustered mapper for different
readmappers. However, we will not pursue that direction in this article.


\section{Results and discussion}
We have selected four datasets to analyze our method.

\begin{itemize}
  \setlength{\itemsep}{1pt}
  \setlength{\parskip}{0pt}
  \setlength{\parsep}{0pt}

\item {\tt SIM}: 2 million simulated 100bp paired-end reads
	(mean insert length 300 and standard
	deviation 30)
	simulated with a coverage of 50x
	obtained with the popular read simulator ART \cite{Huang15022012}
	from human chromosome 17 (from the 10Mbp to 14Mbp region) after
	artificially introducing
	SNPs ({\tt SIM-SNP}) and inserting indels ({\tt SIM-INDEL}).

\item {\tt ECOL}: 21 million 36bp paired-end E. coli reads
  (SRX000429) with a coverage of 160x at a
  genome size of
  5Mb. We estimated the insert length to be 215bp with a standard
  deviation of 10bp.

\item {\tt DROS}: 48 million 76bp paired-end Drosophila reads
  (SRR097732) \cite{Langley01062011} with a coverage of 29x at a
  genome size of
  120Mb. We estimated the insert length to be 320bp with a standard
  deviation of 18bp.

\item {\tt YOR}: 1.46 billion 100bp paired-end Illumina reads of a
  Yoruba individual (ERA015743) (HapMap: NA18507) with a coverage of 46x at a
  genome size of 3.2Gb. We estimated the insert length to be 310bp
  with a standard deviation of 20bp.
\end{itemize}

For all datasets we use $\alpha=\max\{0.5, \frac{31}{L}\}$ and $\beta=0.95$.
The experiments are performed on a Linux machine with 48 AMD
Opteron cores clocked at 2.2 GHz and 256GB
memory. Whenever possible, in particular for multi-threaded tools, 22-24
cores were assigned, and for both multithreaded and non-multi-threaded
programs total system time is reported.

In the following,
we will show the effectiveness
of our clustering algorithm and the benefits of using clusters
in read mapping. We will also discuss how the characteristics
of the read library affect performance and parameter choices.

\subsection{Clustering performance}

Since clustering is a fundamental technique for large data analysis,
many tools have been developed in the past for biological data; i.e.
for protein clustering \cite{Pipenbacher01102002,Loewenstein01072008,Li01072006},
meta-genomics \cite{20388221,10.1371/journal.pone.0003042,10.1371/journal.pbio.0050016,10.1371/journal.pone.0003375},
and expressed sequence tags
(ESTs) \cite{Pertea22032003,Burke01111999,16026600,Hazelhurst15122011}.
Recently, there is a trend of developing alignment-based (for alignment-free clustering see \cite{24011402})
fast clustering tools for very large HTS data sets;
among these, CD-HIT \cite{Fu01122012}, DNACLUST \cite{21718538}, UCLUST \cite{Edgar01102010}
and SEED \cite{Bao:2011ia} are prominent. These four tools
use greedy incremental approaches and apply heuristics to accelerate
clustering (our clustering algorithm also belongs to this group).
Except DNACLUST, which uses a suffix array, these tools also use
some kind of hash/seed based data structure.
Among them, SEED \cite{Bao:2011ia} has been identified as the
state-of-the-art in read clustering. Although it can cluster
up to tens of millions of reads given enough time and memory,
SEED, and other tools, are still not very attractive as a pre-processing
step before read mapping.
There are two reasons behind that: First, they require highly
overlapping reads, usually $\alpha > 0.90$, and, in some cases, high
similarity in the overlap, which results in a large number of clusters;
Second, their
algorithms and data structures are not designed to process billions of
reads.
While large overlap is absolutely necessary for very short reads
(say $< 50$bp), for moderate to large sized reads (say $ > 50$bp),
$\alpha=0.5$ is
sufficient for declaring two reads similar for read mapping purpose
since per base error rate for Illumina reads are small.
To overcome these difficulties, and to support particular
requirements of read mapping, such as paired-end support and sub-optimal
cluster choices, we have designed our own clustering algorithm TreQ-CG.
As a representative of the fast clustering tools (see \cite{Li:2012cq}
for a set of examples) we have selected UCLUST
and SEED, and compared their performance with
TreQ-CG (Table \ref{seed}). We indirectly
measure the quality of the produced
clusters by using them for read mapping in a later step.
Among the two tools in consideration, only SEED allows non-overlapping
bases (at most 6 bases can be excluded), and it allows at
most 3 mismatches in an overlap which corresponds to
parameters
$(\alpha=0.83,\beta=0.90)$,
$(\alpha=0.92,\beta=0.96)$ and
$(\alpha=0.94,\beta=0.97)$ for {\tt ECOL}, {\tt DROS} and
{\tt YOR} respectively.
On the other hand, UCLUST allows a flexible $\beta$ but it does not
allow users to select $\alpha$. Compared to them, TreQ-CG allows both
parameters to be modified. Taking these differences into consideration
we have selected three set of parameters for comparison:
$C_1$ allows UCLUST to run with smallest possible $\beta$ allowed by SEED,
$C_2$ allows smallest possible $\alpha$ and $\beta$ allowed by SEED, and
$C_3$ is the parameter set used by TreQ-CG for subsequent analysis.

Along with memory requirement and running time, our primary criteria for
evaluating clustering tools is the number of clusters produced, which
is normalized as $\mbox{\% of clusters} = \frac{\mbox{\# of anchor reads}}{\mbox{\# of reads}}$.
Since SEED depends on large overlap between reads, it produces $37.79\%$
clusters for {\tt DROS} in $7.93$ hours and does not complete on the
{\tt YOR} dataset. On the other hand, UCLUST's use of full overlap
and gapped alignment makes it very slow on large datasets.
It takes 52 hours on the {\tt ECOL} dataset but on the larger datasets
it did not complete in one week. In comparison,
TreQ-CG
works with a wide
range of parameter choices in both single-end and paired-end mode,
produces fewer clusters, and runs faster than SEED.
Additionally, for parameter set $C_3$, TreQ-CG discards
$0.3\%$, $1.6\%$, and
$5.2\%$ reads as bad reads in single-end mode, and $0.2\%$, $1.3\%$, and
$4.8\%$ reads in paired-end mode.
The clusters
produced by SEED and other clustering tools on smaller datasets such as
{\tt ECOL} are useful for other applications, for example in meta-genomics.
But unless they support smaller overlap length, which is difficult for their
particular choice of data structures and algorithms, they will not be
attractive choices for pre-processing reads before read mapping.
We do not consider them for subsequent analysis in the following section.

    \begin{table}
    \centering
    \caption{
    The performance of three clustering tools, UCLUST, SEED, and TreQ-CG,
    on three biological datasets {\tt ECOL}, {\tt DROS}, and {\tt YOR}
    is shown. Three different parameter configurations are used for
    comparison.
    SEED was run with parameters {\tt --shift X --mismatch Y --fast --reverse},
    where {\tt X} and {\tt Y} corresponds to $\alpha$ and $\beta$ respectively in TreQ-CG.
    Here, single-end and paired-end are abbreviated as S.E. and P.E..
    \label{seed}
    }

  \begin{tabular}{lllrrr}
  \toprule
   \multirow{2}{*}{Dataset} &
   \multirow{1}{*}{Parameters} &
   \multirow{2}{*}{Program} &
   Time & Memory & Clusters\\
                               &$(\alpha,\,\beta)$&& (h:mm) &   (GB) &            ($\%$) \\
   \midrule
   \multirow{12}{*}{ {\tt ECOL} }
	      & \multirow{4}{*}{\begin{tabular}{l}$C_1$\\$1.00, 0.90$\end{tabular}} &  UCLUST & 52:07 &  {\bf 2} & 17.47 \\ 
						    &&    SEED &  0:32 & 12 & 38.97 \\ 
						    && TreQ-CG (S.E.)&  0:24 & 17 & 23.86 \\ 
						    && TreQ-CG (P.E.)&  0:30  & 17 & 31.21 \\ 
   \cline{2-6}
              & \multirow{4}{*}{\begin{tabular}{l}$C_2$\\$0.83, 0.90$\end{tabular}} &  UCLUST &    --- & --- & --- \\ 
						    &&    SEED & 0:27 & 11 & 5.71 \\
						    && TreQ-CG (S.E.) & {\bf 0:12} & 16 & {\bf 4.07} \\
						    && TreQ-CG (P.E.) & {0:13} & {16} & {5.14} \\
   \cline{2-6}
              & \multirow{4}{*}{\begin{tabular}{l}$C_3$\\$0.86, 0.95$\end{tabular}} &  UCLUST &    --- & --- & --- \\ 
						    &&    SEED & 0:30 & 11 & 7:30 \\ 
						    && TreQ-CG (S.E.)& 0:14 & 16 & 7.30 \\
						    && TreQ-CG (P.E.)& 0:15 &  16 & 8.57 \\
   \hline
   \multirow{11}{*}{ {\tt DROS} }
	      & \multirow{4}{*}{\begin{tabular}{l}$C_1$\\$1.00, 0.96$\end{tabular}} 
						    &  UCLUST &    --- & --- & --- \\ 
						    &&    SEED & 3:22 & 25 &   75.97 \\ 
						    && TreQ-CG (S.E.)& 3:36 & 34 &   70.34 \\
						    && TreQ-CG (P.E.)&  4:02  & 35 & 76.73 \\
   \cline{2-6}
              & \multirow{4}{*}{\begin{tabular}{l}$C_2$\\$0.92, 0.96$\end{tabular}} &  UCLUST &    --- & --- & --- \\
						    &&    SEED & 7:56 & 25 & 37.79 \\
						    && TreQ-CG (S.E.)& 2:00 & 23 & 24.50 \\
						    && TreQ-CG (P.E.)& 2:30 &  24 & 30.09 \\
   \cline{2-6}
              & \multirow{3}{*}{\begin{tabular}{l}$C_3$\\$0.50, 0.95$\end{tabular}} &  UCLUST, SEED &    --- & --- & --- \\
						     && TreQ-CG (S.E.)& {\bf 1:09} & {\bf 20}  & {\bf 7.29} \\
						     && TreQ-CG (P.E.) &  { 1:18} &  {20} & {9.18} \\
   \hline
   \multirow{8}{*}{ {\tt YOR} }
	      & \multirow{2}{*}{\begin{tabular}{l}$C_1$\\$1.00, 0.97$\end{tabular}} 
						    &  UCLUST, SEED &    \multirow{2}{*}{---} & \multirow{2}{*}{---} & \multirow{2}{*}{---} \\
						    &&  TreQ-CG & & & \\
   \cline{2-6}
              & \multirow{3}{*}{\begin{tabular}{l}$C_2$\\$0.94, 0.97$\end{tabular}} 
						    &  UCLUST, SEED &    --- & --- & --- \\
		                                     && TreQ-CG (S.E.)&    197:35 &   196 & 29.45 \\
		                                     && TreQ-CG (P.E.)&    --- &   --- & --- \\
   \cline{2-6}
	      & \multirow{3}{*}{\begin{tabular}{l}$C_3$\\$0.50, 0.95$\end{tabular}}
						    &  UCLUST, SEED &    --- & --- & --- \\
						    && TreQ-CG (S.E.)& {\bf 107:52} &  {\bf 76} & {\bf 7.48} \\
						    && TreQ-CG (P.E.)& {133:22} & { 91} & { 10.90} \\
  \bottomrule
  \end{tabular}    
    \vspace{-6mm}
  \end{table}

\subsection{Read mapping performance}
\label{cluster-tool-comp}
On biological datasets there is no ground truth
available for deciding whether a read is mapped correctly.  There are
also differences---in particular for low-quality reads and in presence
of genetic variations---in mapping performance between individual readmappers, and
likewise we expect some differences between clustered mapping and
individual mapping even with the same readmapper.

We compare clustered and individual mapping for Bowtie 2
\cite{Langmead2012} version 2.1.10, BWA \cite{Li15072009} version 0.5.9,
Novoalign (\url{http://www.novocraft.com/}) version 2.07.13 and
Stampy \cite{Lunter2011} version 1.0.19.
All mappers, chosen because they are widely used and implement different
algorithmic approaches, were run with default parameters except forcing
reporting of a single best hit. We expect other readmappers, specially highly sensitive
computationally intensive ones, to benefit from clustered approach and their overall performance to
follow the performance of the readmappers chosen for comparison.

Including clustering time, clustered read mapping achieves a speed-up
compared to the time of mapping individual reads for all readmappers
tested, see Table~\ref{runtime-se} for single-end reads and supplementary Table S.1 for paired-end reads. The speed-up ranges from
8.9 using Stampy on {\tt YOR} to 1.4 using Novoalign on {\tt ECOL}.
One would expect that the time needed for mapping anchor reads should
be proportional to the number of anchors. But in practice, reads
with
high sequencing errors form singleton clusters (that
is a cluster without any member except the anchor read) and read
mappers take more time to map these reads.  As a result the speed-up is
not exactly proportional to the number of clusters. Still, due to the achieved
speed-up, the use of sensitive readmappers
like Stampy becomes computationally as feasible as running BWA.

For qualitative assessment of clustered read mapping we use the following
three metrics.

\begin{itemize}
  \setlength{\itemsep}{1pt}
  \setlength{\parskip}{0pt}
  \setlength{\parsep}{0pt}
\item {\bf Accuracy.} Since the true originating
position of a read
is known for simulated data, we define accuracy as the proportion of reads that
map within $\pm L$ bases of the true position \cite{Hamada:2011bg}. We do
not require a
strict match because there can be
ambiguity in the starting position of the optimal alignment
due to the presence of gaps and exact alignment parameters.

\item {\bf Alternate mapping rate.} Since the ground truth is not known for biological data, we compute alternate mapping rate between individual mapping and clustered mapping. Alternate mapping rate is defined as the percentage of reads for which either both approaches
report mapping positions not within $\pm L$ bases of each other, or one approach maps the read while the other fails. 
This essentially measures how well the clustered read mapping recreates the {\em exact} output of the individual read mapping.

\item {\bf Concordance.} For biological data, another
important measure is concordance of read pairs. We define it as the
proportion of read pairs for which the estimated insert size is within
the mean insert length of the library, allowing for $\pm 5$ standard
deviation. It has been argued in the literature \cite{Lunter2011} that
the concordance of paired-end reads mapped in single-end mode provides an
indirect measure of mapping accuracy.
\end{itemize}

\begin{figure*}
\centering
\includegraphics[scale=0.45]{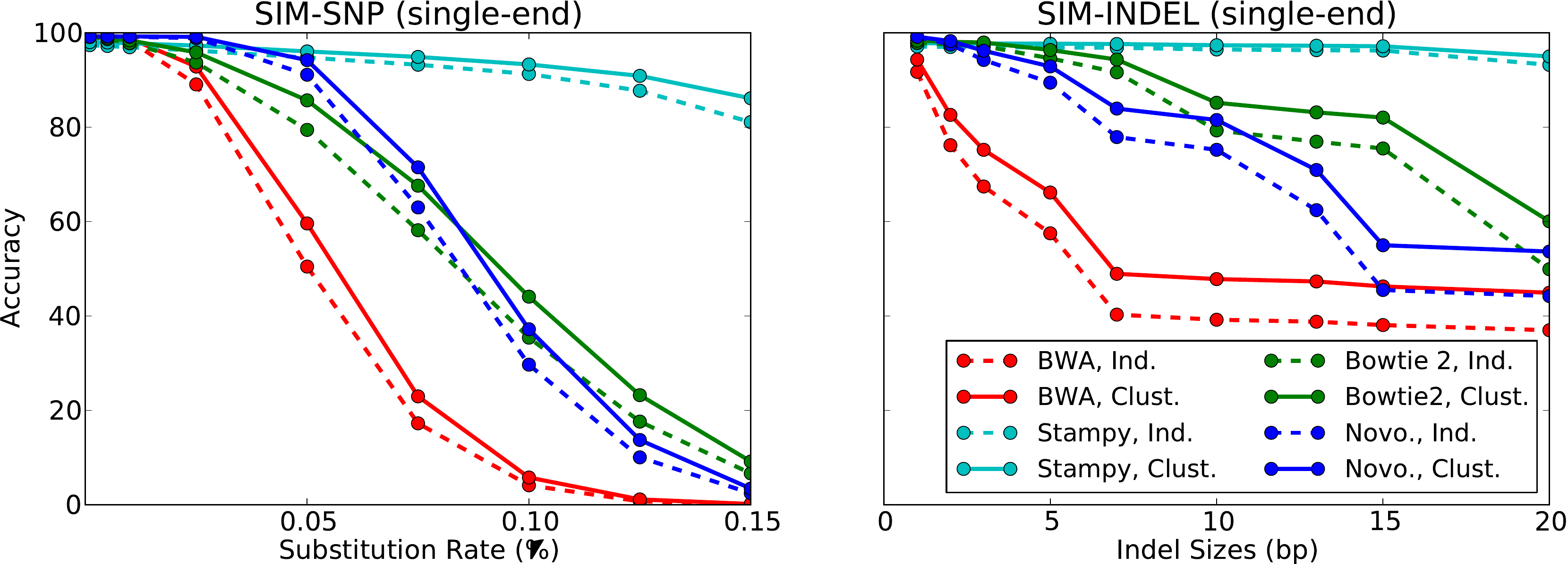}
\caption{\label{fig:sim_res_se}
{\bf Quality improvements in detecting SNPs and structural variants.}
    Accuracy of mapping single-end simulated reads between individual and clustered approach
    is shown. As expected, BWA and Bowtie 2 do not perform very well in presence
    of large variations. 
    Clustered read mapping performs significantly better
    than individual BWA and Bowtie 2, and generally agrees with the individual version
    of the sensitive readmappers.
}
\end{figure*}

\begin{figure*}
  \centering
  \includegraphics[scale=0.5]{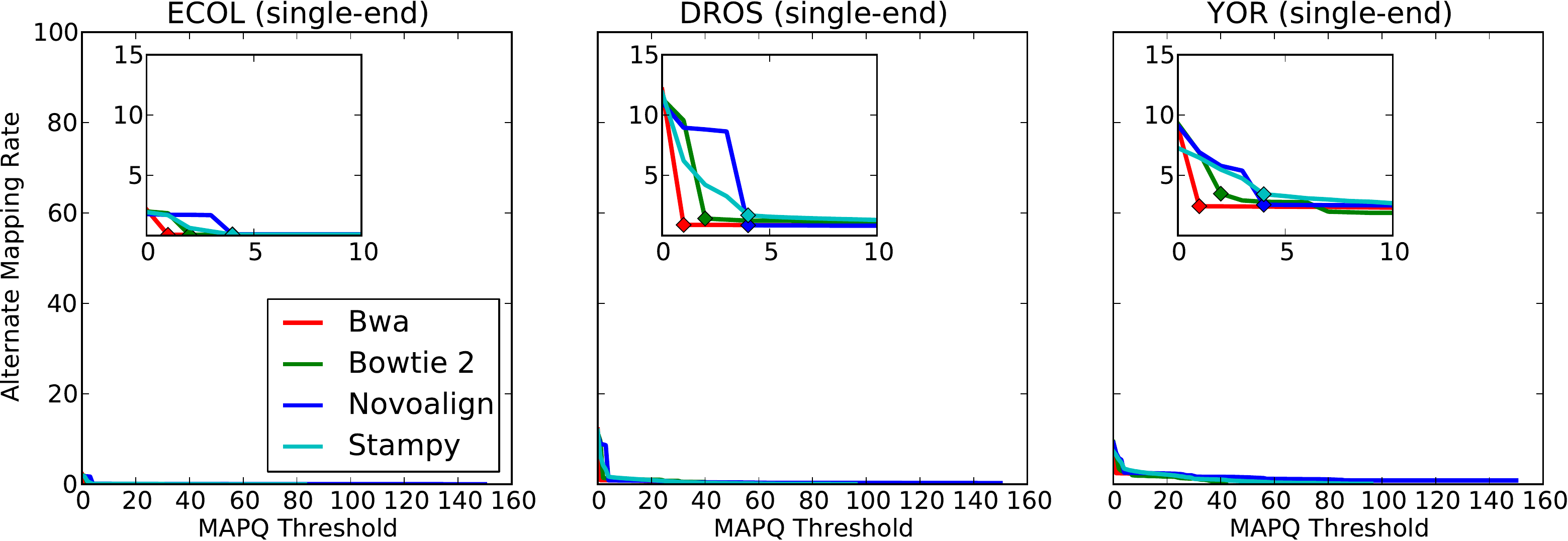}
  \caption{ \label{fig:real-se}
  {\bf Alternate mapping rate between clustered and individual single-end read mapping.} Alternate mapping rate is shown as a function of MAPQ threshold. Maximum reported MAPQ value varies between readmappers; for BWA, Bowtie 2, Novoalign and Stampy maximum reported MAPQ for single-end reads are respectively 37, 42, 150 and 96. The cutoffs used to report alternate mapping rate in Table~\ref{runtime-se} are shown with diamond signs in the inset figures.}
  \vspace{-4mm}
\end{figure*}

\subsubsection{Simulated Data}
We have tested the effect of SNPs and indels on single-end and
paired-end clustered mapping; the results are reported in
Figure~\ref{fig:sim_res_se} (single-end) and S.1 (paired-end). For {\tt SIM-SNP} we have modified the sample
genome with substitution rate ranging from 0.001 to 0.15 while allowing
no indels. For {\tt SIM-Indel} we have inserted indels of a particular size
$I\in [1, 20]$ with an indel rate of 0.01 per base.
For each SNP and indel size configuration we have sequenced 2 million
reads using the read simulator ART, which applies Illumina-specific
sequencing errors on top of the genetic variations present in the
artificial sample genome.
Since indels are randomly placed in the sample genome
before sequencing, approximately 40\% reads in the {\tt SIM-Indel}
dataset do not contain any indel. As expected, Stampy's performance stays
excellent for the range of variations tested but others do not perform well
in presence of high genetic variation. In case of Stampy,
clustered mapping closely follows individual mapping.
On the other hand, clustered mapping performs better
than individual BWA, Bowtie 2, and Novoalign.
To gain efficiency these readmappers limit their mapping process through
limited edit distance or score cutoff.
In contrast, clustered mapping is more permissive due to low
score cutoff (default $\frac{L}{3}$).

\subsubsection{Biological Data}
It is reasonable to expect that the reads mapped with low alignment quality are those which should cause the most differences between clustered and individual mapping, and consequently influence the accuracy measures. We use the MAPQ value reported in the SAM file, where higher value indicates higher confidence in a particular alignment, as thresholds to limit the read alignments considered to compute alternate mapping rate. In practice, along with unmapped reads (reported with default MAPQ value 0), low quality alignments are frequently excluded from downstream analysis.  Following~\cite{Li15072009}, we select MAPQ value of 1 and 2 for BWA and Bowtie 2 respectively, and 4 for Novoalign and Stampy, as a cutoff to report alternate mapping rate in Table~\ref{runtime-se} and Table S.1. Alternate mapping rates computed using different values of MAPQ threshold are reported in Fig.~\ref{fig:real-se} and Fig. S.2. We also report concordance of reads mapped in Table~\ref{runtime-se} and Table S.1.

From Figure~\ref{fig:real-se} we observe that the alternate mapping rate among all reads above a MAPQ threshold strictly decreases as a function of the threshold and eventually drops below 1\%. 
At the selected MAPQ thresholds, depending on the readmapper used, clustered mapping reports alternate mapping for about 0--1\%, 1--3\%, and 2--4\%  reads for {\tt ECOL}, {\tt DROS} and {\tt YOR} datasets respectively. Compared to single-end reads, there is an 1-2\% increase in alternate mapping rate for paired-end mode (see Table S.1). The observations are consistent with alternate mapping rates when comparing two different readmappers and are largely explained by differences in alignment parameters (see Fig. S.4).
As for concordance of single-end reads in
clustered approach, there is a small gain in concordance  for {\tt ECOL} and {\tt DROS}, and a small loss in concordance for {\tt YOR} (Table~\ref{runtime-se}). On the other hand, for paired-end reads, except for {\tt ECOL} dataset, there is always a gain in concordance, which is mostly due to altering mapping locations based on pairing information (see Table S.1). Increase in alternate mapping rate and gain in concordance for paired-end reads indicates that our paired-end mapping might be too permissive compared to what is allowed by an individual readmapper.

\begin{table*}
  \centering
    \caption{
    Running time and memory requirement of individual and clustered read mapping on single-end datasets are reported.
    As expected, clustered read mapping achieves significant speed-up for all datasets. Although
    running Stampy on {\tt YOR} dataset in individual mapping mode is prohibitively expensive,
    clustered approach makes running Stampy as feasible as running BWA. \label{runtime-se}}
    \footnotesize
\begin{tabular}{llcrrrrrccrr}
\toprule
\multirow{3}{*}{\begin{tabular}{l}Dataset\\(single-\\end)\end{tabular}} & \multirow{3}{*}{Mapper} & \multicolumn{3}{c}{Memory (GB)}&\multicolumn{3}{c}{Time (hh:mm)}& \multirow{3}{*}{Speed-up}&\multicolumn{3}{c}{Mapping quality} \\
\cmidrule{3-8}\cmidrule{10-12}
			    & & \multirow{2}{*}{\begin{tabular}{l}Individual\\ mapping\end{tabular}} & \multicolumn{2}{c}{TreQ-CG}
			    & \multirow{2}{*}{\begin{tabular}{l}Individual\\ mapping\end{tabular}} & \multicolumn{2}{c}{TreQ-CG} & & \multirow{2}{*}{\begin{tabular}{l}Alternate\\ mapping rate\end{tabular}}&\multicolumn{2}{c}{Concordance}\\
\cmidrule{4-5}\cmidrule{7-8}\cmidrule{11-12}
			    & &  & Clust. & Map. & & Clust. & Map. & & & Ind. & Clust.\\
\midrule
\multirow{4}{*}{\tt ECOL}
& \multirow{1}{*}{BWA} 	   	& {\bf 0.1} &\multirow{4}{*}{16}& 0.9 & 0:49 &\multirow{4}{*}{0:14}& {\bf 0:12} & 1.9 & 0.06 & 92.93 & {\bf 93.44}\\
& \multirow{1}{*}{Bowtie~2} 	& {\bf 0.3} && 0.9 & 0:43 && {\bf 0:13} & 1.6 & 0.06 & 93.41 & {\bf 93.64}\\
& \multirow{1}{*}{Novoalign}	& {\bf 0.0} && 1.0 & 0:38 && {\bf 0:14} & 1.4 & 0.10 & 96.59 & {\bf 96.80}\\
& \multirow{1}{*}{Stampy}   	& {\bf 0.0} && 1.0 & 4:17 && {\bf 0:28} & 6.1 & 0.08 & {\bf 96.23} & 96.19\\
\hline
\multirow{4}{*}{\tt DROS}
& \multirow{1}{*}{BWA} 		& {\bf 0.3} &\multirow{4}{*}{20}& 1.4 &  3:29 &\multirow{4}{*}{1:09}& {\bf 0:49} & 1.8 & 0.88 & 71.02 & {\bf 72.07}\\
& \multirow{1}{*}{Bowtie~2}	& {\bf 0.4} && 1.4 & 2:53 && {\bf 0:39} & 1.6 & 1.41 & {\bf 74.82} & 73.87\\
& \multirow{1}{*}{Novoalign} 	& {\bf 0.6} && 1.4 & 6:19 && {\bf 1:09} & 2.8 & 0.85 & 71.20 & {\bf 73.87}\\
& \multirow{1}{*}{Stampy}    	& {\bf 0.3} && 1.4 &28:08 && {\bf 2:59} & 6.8 & 1.68 & 74.06 & {\bf 74.31}\\
\hline
\multirow{4}{*}{\tt YOR}
& \multirow{1}{*}{BWA}       	& {\bf 5.0} &\multirow{4}{*}{76}&  8.3 & 477:30 &\multirow{4}{*}{107:52}& {\bf 157:29} & 1.8 & 2.43 & {\bf 78.43} & 77.54\\
& \multirow{1}{*}{Bowtie~2}   	& {\bf 3.5} && 8.2 & 307:45 && {\bf 75:04}  &  1.7 & 3.47 & {\bf 82.38} & 79.75\\
& \multirow{1}{*}{Novoalign} 	& {\bf 7.8} && 8.3 & 714:07 && {\bf 230:47} &  2.1 & 2.55 & {\bf 79.40} & 78.24\\
& \multirow{1}{*}{Stampy}    	& {\bf 2.7} && 8.2 & 3203:39&& {\bf 251:40} &  8.9 & 3.43 & {\bf 80.51} & 78.87\\
\bottomrule
\end{tabular}
\vspace{-5mm}
\end{table*}

\section{Conclusions}
We propose to cluster high-throughput sequencing reads as a first step
in the analysis and show,
as a proof of concept,
that our greedy clustering algorithm
accelerates
read mapping.
We observe speed-ups for all readmappers tested ranging from 1.4 to 8.9
with
very little alternate mapping
between individual and clustered mapping for
high-quality reads, and
low level of alternate mapping rate
for all reads with a small loss
in concordance for some datasets.
We do expect even more favorable results
for
read mapping if there is a lot of genetic variation in the sample
genome, such as in cancer datasets, or the evolutionary distance
between sample and reference genome is large.

There are several areas in which the method can
be improved, the most important ones being cluster quality and
size, as running time is inversely proportional to
the total number of clusters.
It seems reasonable that the
singleton clusters (that is a cluster with cardinality one, which arises either due to high sequencing error or very low coverage)
can be reduced by
additional passes, possibly with a smaller value of $k$ for
k-mers. The additional effort in clustering is very likely to be
made up
in mapping, by net savings in running time.
So far read mapping is performed with anchor reads without any
post-processing.
An obvious improvement can be made by considering
clusters as an instance of multiple sequence alignment problem,
and correcting errors using the alignment information.
Moreover, computing an elongated consensus sequence and using it
instead of the anchor read is also desirable.
However, one has to judiciously choose, based on coverage
and within cluster sequence variation, the actual fragment of the
consensus to use for mapping to see further improvements.


\bibliographystyle{abbrv}
\bibliography{combined}

\newpage

\setcounter{figure}{0} 
\renewcommand{\thefigure}{S.\arabic{figure}}
\setcounter{table}{0} 
\renewcommand{\thetable}{S.\arabic{table}}

\section*{Appendix}

\subsection*{Results for paired-end datasets}

\begin{figure*}
\centering
\includegraphics[scale=0.45]{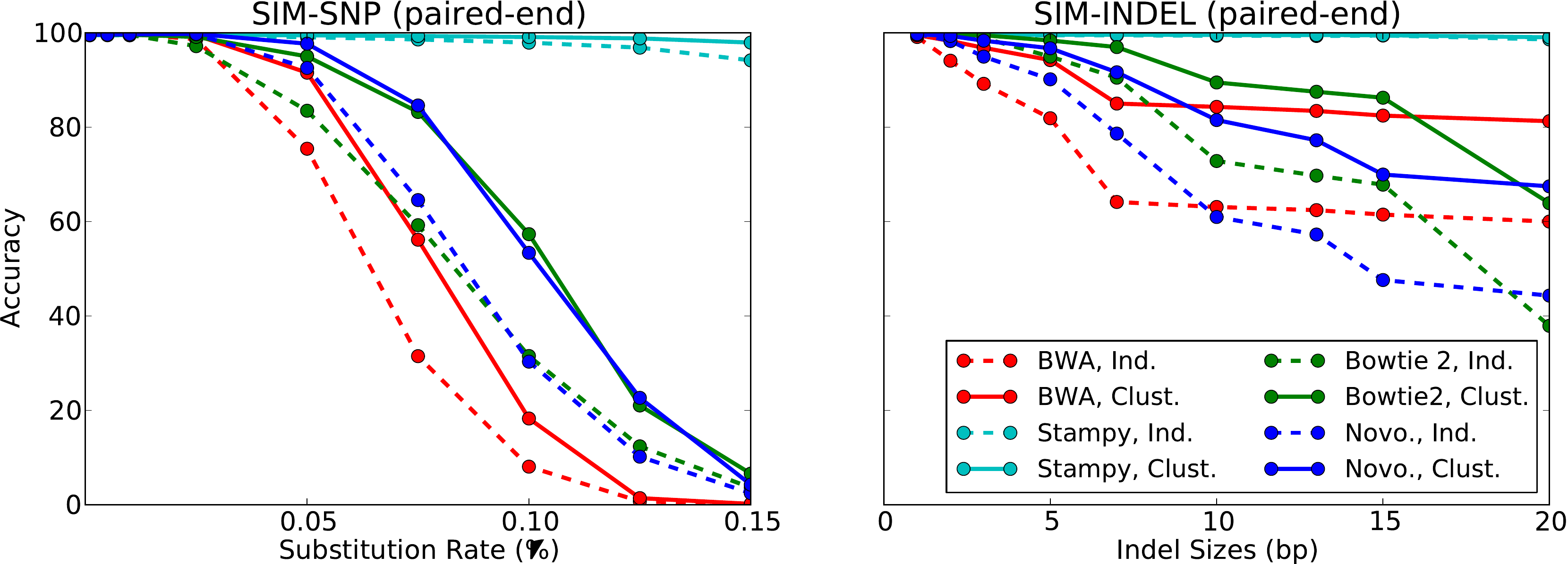}
\caption{\label{fig:sim_res_pe}
{Quality improvements in detecting SNPs and structural variants.}    }
\end{figure*}

\begin{figure*}
  \centering
  \includegraphics[scale=0.5]{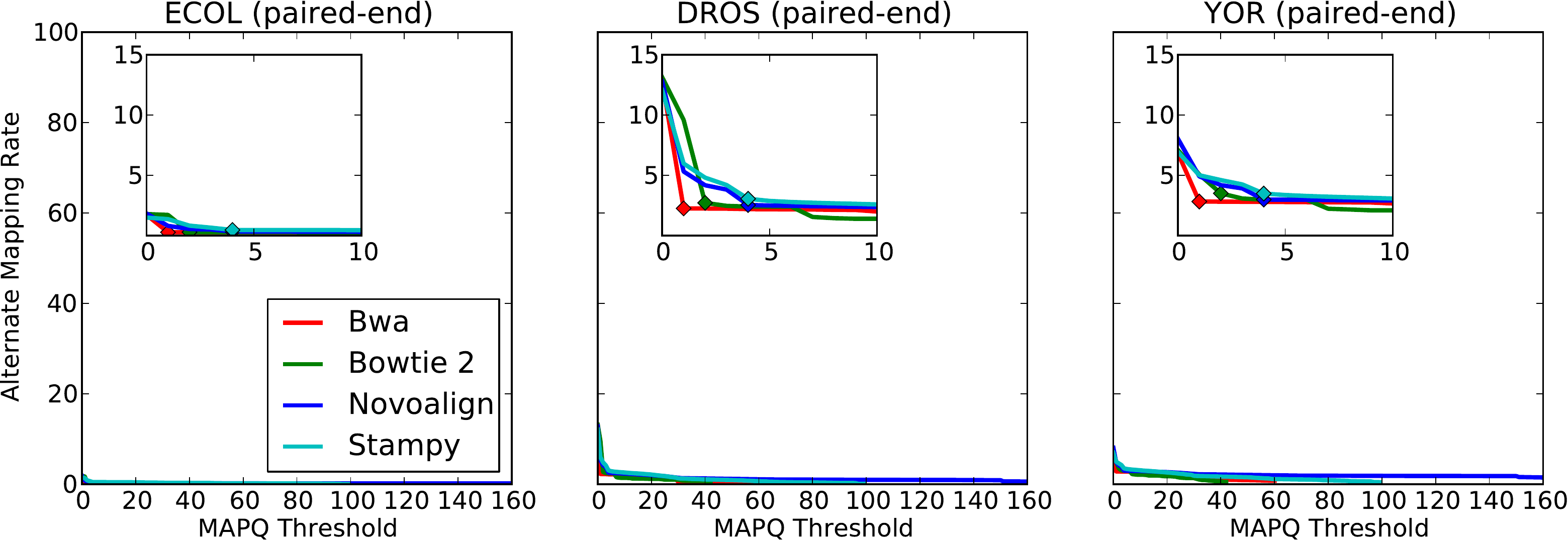} 
  \caption{ \label{fig:real}
  {\bf Alternate mapping rate between clustered and individual paired-end read mapping.} Alternate mapping rate is shown as a function of MAPQ threshold. Maximum reported MAPQ value varies between readmappers; for BWA, Bowtie 2, Novoalign and Stampy maximum reported MAPQ for single-end reads are respectively 60, 42, 159 and 99. The cutoffs used to report alternate mapping rate in Table~\ref{runtime-pe} are shown with diamond signs in the inset figures.}
\end{figure*}

Accuracy of mapping simulated paired-end reads between individual and clustered approach is shown in Fig.~\ref{fig:sim_res_pe}. Running time, memory requirement and mapping quality of individual and clustered paired-end read mapping are reported in Table~\ref{runtime-pe}. Read mapping agreement is shown in Fig.~\ref{fig:real} as a function of MAPQ threshold.    

\begin{table*}
\centering
    \caption{ { Running time and memory requirement of paired-end clustered read mapping.}     
    \label{runtime-pe}} 
    \footnotesize
\begin{tabular}{llcrrrrrccrr}
\toprule
\multirow{3}{*}{\begin{tabular}{l}Dataset\\(paired-\\end)\end{tabular}} & \multirow{3}{*}{Mapper} & \multicolumn{3}{c}{Memory (GB)}&\multicolumn{3}{c}{Time (hh:mm)}& \multirow{3}{*}{Speed-up}&\multicolumn{3}{c}{Mapping quality} \\
\cmidrule{3-8}\cmidrule{10-12}
			    & & \multirow{2}{*}{\begin{tabular}{l}Individual\\ mapping\end{tabular}} & \multicolumn{2}{c}{TreQ-CG}
			    & \multirow{2}{*}{\begin{tabular}{l}Individual\\ mapping\end{tabular}} & \multicolumn{2}{c}{TreQ-CG} & & \multirow{2}{*}{\begin{tabular}{l}Alternate\\ mapping rate\end{tabular}}&\multicolumn{2}{c}{Concordance}\\
\cmidrule{4-5}\cmidrule{7-8}\cmidrule{11-12}			    
			    & &  & Clust. & Map. & & Clust. & Map. & & & Ind. & Clust.\\			    
\midrule 
\multirow{4}{*}{\tt ECOL}
& \multirow{1}{*}{BWA} 	   	& {\bf 0.2} &\multirow{4}{*}{16}& 0.9 & 1:01 &\multirow{4}{*}{0:15}& {\bf 0:13} & 2.2 & 0.27 & {\bf 98.77} & 98.62\\
& \multirow{1}{*}{Bowtie~2} 	& {\bf 0.3} && 0.9 & 0:49 && {\bf 0:12} & 1.8 & 0.28 & {\bf 98.39} & 98.36\\
& \multirow{1}{*}{Novoalign}	& {\bf 0.0} && 1.0 & 0:57 && {\bf 0:15} & 1.9 & 0.33 & {\bf 99.57} & 99.28\\
& \multirow{1}{*}{Stampy}   	& {\bf 0.0} && 1.0 & 4:08 && {\bf 0:31} & 5.4 & 0.33 & {\bf 99.41} & 99.18\\
\hline
\multirow{4}{*}{\tt DROS} 
& \multirow{1}{*}{BWA} 		& {\bf 0.4} &\multirow{4}{*}{20}& 1.4 &  3:57 &\multirow{4}{*}{1:18}& {\bf 1:09} & 1.6 & 2.25 & 84.17 & {\bf 85.62} \\
& \multirow{1}{*}{Bowtie~2}	& {\bf 0.5} && 1.4 &  5:33 && {\bf 1:05} & 2.3 & 2.71 & 84.56 & {\bf 85.51} \\
& \multirow{1}{*}{Novoalign} 	& {\bf 0.6} && 1.4 & 11:25 && {\bf 1:35} & 4.0 & 2.53 & 82.82 & {\bf 85.24} \\
& \multirow{1}{*}{Stampy}    	& {\bf 0.3} && 1.4 & 38:51 && {\bf 4:17} & 7.0 & 3.05 & 82.52 & {\bf 86.08}\\
\hline	
\multirow{4}{*}{\tt YOR}
& \multirow{1}{*}{BWA}       	& {\bf 5.0} &\multirow{4}{*}{91}& 11.0 & 533:38 &\multirow{4}{*}{133:22}& {\bf 171:31} & 1.8 & 2.82 & 89.24 & {\bf 90.56}\\
& \multirow{1}{*}{Bowtie~2}   	& {\bf 3.7} && 10.9 &  352:49 &&  {\bf 89:11} & 1.6 & 3.50 & 90.48 & {\bf 91.46}\\
& \multirow{1}{*}{Novoalign} 	& {\bf 8.0} && 10.9 &  969:19 && {\bf 368:39} & 1.9 & 2.98 & 87.52 & {\bf 90.41}\\
& \multirow{1}{*}{Stampy}    	& {\bf 2.7} && 10.9 & 3397:02 && {\bf 453:59} & 5.8 & 3.48 & 88.38 & {\bf 91.18}\\
\bottomrule 
\end{tabular}
    \end{table*}

\subsection*{Additional points}

\subsubsection*{Coverage and read length} 
Statistically significant
findings in the downstream analysis require high coverage sequencing.
For example, Illumina's cancer pipeline recommends 40x coverage for
normal and 80x coverage for cancer tissues. The higher the coverage of
the library the more advantageous the clustering becomes, as we expect
the average cluster size to increase.  We tested read libraries with
reads of 36--100bp and we expect results to improve for even longer
reads. For significantly shorter reads read mapping based on clusters
will likely fail to achieve any significant speed-up or 
the loss of sensitivity will be non-negligible.

\subsubsection*{Bias of 15-mer} 
An overlap of length at least $15\cdot(E+1)+E$ 
is guaranteed to overcome $E$ errors in our algorithm. Since we use a
fixed value for $\alpha$, some reads will not be detected as members and
possibly create singleton clusters. Smaller $k$ will lessen this problem,
but to keep the algorithm efficient for large datasets we do not want
to lower $k$. Instead, if a read library is suspected to contain many
reads with large number of errors $\alpha$ can be increased to deal with
the issue.

\subsubsection*{Parameter choices} 
\label{sec:params}
If the sequencing error rate $\epsilon$
(typically less than $0.02$ for Illumina experiments) can be determined 
\emph{a priori}, 
we suggest using a similarity cutoff $\beta \le (1-2\epsilon)$. Too 
large a value for $\beta$ makes it difficult to assign a read to clusters, 
thus creating too many clusters,
while a value too low might assign reads to wrong clusters. 

For $\alpha$, if we choose a smaller value, we will get fewer clusters,
which will in turn decrease running time, but unfortunately read mapping 
sensitivity will drop. On the other hand, a value too large leads to more
clusters. This increases running time of both clustering and mapping,
but it also increases mapping sensitivity. The values we used,
$\alpha=\max\{0.5, \frac{31}{L}\}$ and $\beta=0.95$, worked well for 
three genomes of very
different sizes and we expect that the choice will work well for
experiments with similar coverage, read length and genome size and
structure. Should the parameters of the experiment change drastically
the choices might have to be re-evaluated in a preliminary study.

\subsubsection*{Automatic choice of parameters} It is possible to design
a scheme to automatically select the necessary parameters $\alpha$ and
$\beta$.  
A small, randomly selected set of reads can be mapped to the reference 
genome to compute the error rate $\epsilon$ from the uniquely mapped reads.
Given $\epsilon$, we can perform a grid search over the possible
values of $\alpha$ and $\beta$. 
A chosen set of parameters should fulfill
two criteria: One, a valid overlap, defined by $\alpha$ and $\beta$,
between two reads from different genomic locus
should only occur with very low probability, 
and two, the amount of required memory,
which is primarily determined by the number of clusters, should be less
than the available system memory. Next, we show an analysis of the 
expected number of clusters produced by a set of parameters. 

\paragraph{Number of clusters:}
We will assume that the reads are numbered in the order they are processed.
Let $r'_i \ne r'_j$ means that read $r_i$ has an overlap with read $r_j$ of 
length at least $\alpha L$ and their overlapped portion $r'_i$ and $r'_j$ 
has similarity less than $\beta$. 
Let the number of sequencing errors in $r'_i$ be $s(r'_i)$. Assuming
$s(r'_i)$ is binomially distributed with rate $\epsilon$, 
$Pr[ s(r'_i) = k ] = {|r'_i| \choose k} \epsilon^k (1-\epsilon)^{|r'_i|-k}$.
Let $C_i \subseteq \{r_1, r_2, \cdots , r_{i-1}\}$ be the set of 
anchor reads with at least $\alpha L$ overlap with $r_i$. 
Given $C_i$, the probabity of read $r_i$ becoming an anchor read and 
forming it's own cluster is 
$P_i = Pr\big[ \bigcap\limits_{r_j \in C_i} r_i' \ne r_j' \big]$.
We simplify the analysis by assuming that all the reads in 
$C_i$ has overlap of length $l'=\frac{(1+\alpha)}{2} L$, which is the 
expected length of a valid overlap. The updated probability is 
$P_i^* \approx P_i$.
Since the number of mismatches allowed is at most  
$ m =  (1-\beta)l'$, $P_i^*$ is upper bounded by
\begin{align*}
& \sum\limits_{x=0}^{l'} \bigg( Pr[s(r'_i)=x] \prod_{r_j \in C_i} Pr[s(r'_j) \ge m - x] \bigg) \\
&= \sum\limits_{x=0}^{l'} Pr[s(r'_i)=x] Pr[s(r'_j) \ge m - x]^{|C_i|} \\
&= \sum\limits_{x=0}^{m - 1} Pr[s(r'_i)=x] Pr[s(r'_j) \ge m - x]^{|C_i|} + Pr[s(r'_i) \ge m] .
\end{align*}
Let $T_{i-1}$ be the number of clusters formed after processing $i-1$ reads.
Assuming the reads were sequenced uniform randomly from the genome,
$E[|C_i|] = \sum\limits_{j=0}^{i-1} \frac{2(1-\alpha)L}{G} P_{j} = \frac{2(1-\alpha)L}{G} T_{i-1}$.
Now, if we replace $|C_i|$ with $E[|C_i|]$, $P_i^*$ is approximately equal to
\begin{align*}
& \approx \sum\limits_{x=0}^{m - 1} Pr[s(r'_i)=x] 
Pr[s(r'_j) \ge m - x]^{\frac{2(1-\alpha)L}{G} T_{i-1}} + Pr[s(r'_i) \ge m] \\
& = \sum\limits_{x=0}^{m - 1} 
{ l' \choose x} \epsilon^x (1-\epsilon)^{l'-x} 
\Bigg( \sum\limits_{y = m-x}^{l'} { l' \choose y} \epsilon^{y} (1-\epsilon)^{l'-y}\Bigg)^{\frac{2(1-\alpha)L}{G} T_{i-1}} \\
& \hspace{50mm} + \sum\limits_{y = m}^{l'} { l' \choose y} \epsilon^{y} (1-\epsilon)^{l'-y}.
\end{align*}

\begin{figure*}
\begin{center}
\includegraphics[scale=0.5]{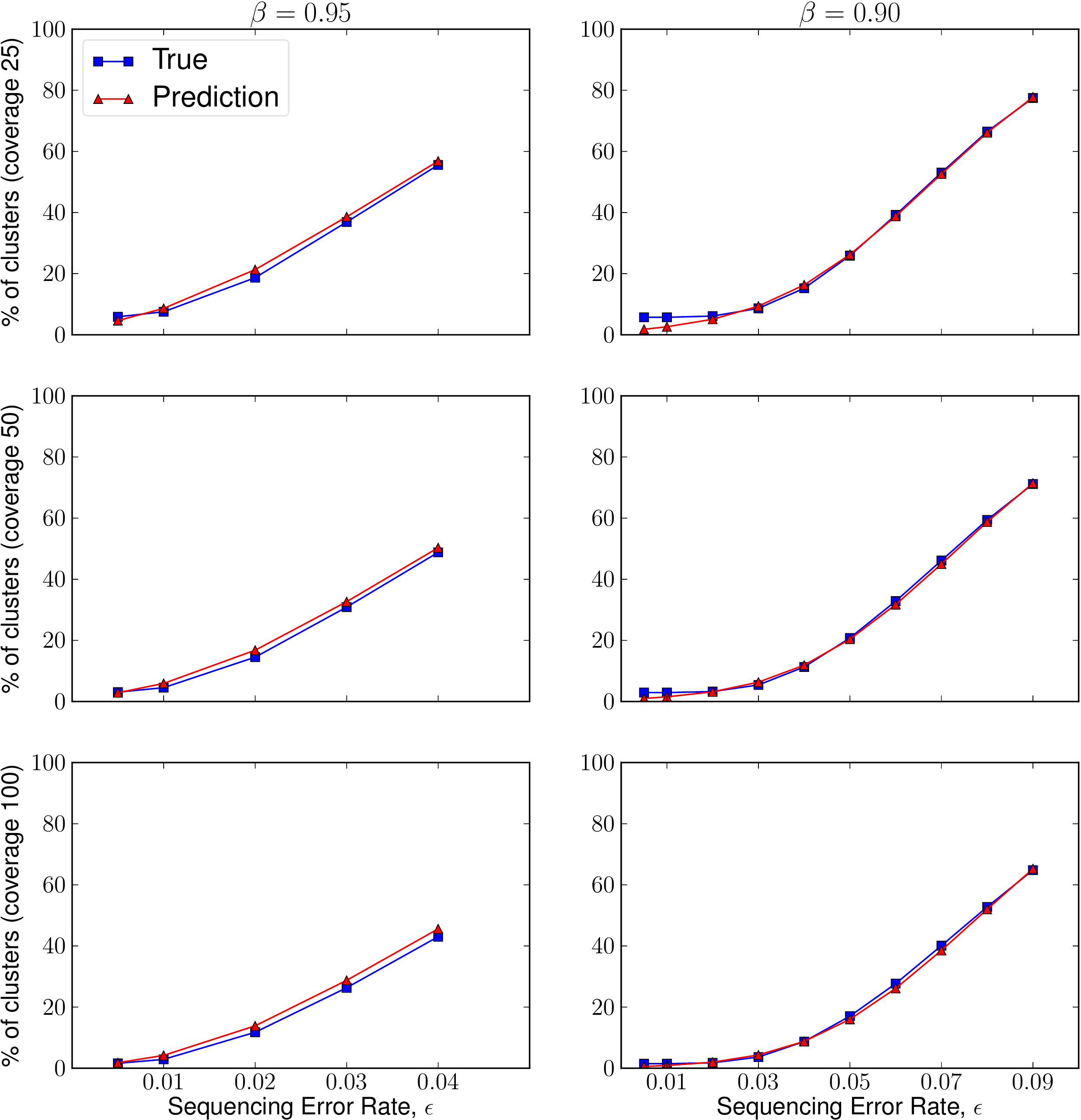}
\end{center}
\caption{
{\bf Expected number of clusters.}
From an artificially generated sample genome,
an i.i.d. sequence of 1 million bases,
we sequence 100bp single-end reads with coverage 25, 50, and 100, and insert
sequencing error at a rate ranging from 0.005 to 0.09. 
We compare the true number of clusters computed using the clustering algorithm
to the numerically evaluated ones
for $\beta=0.95$ and $\beta=0.9$.
The total number of cluster is computed very
accurately except for small values of $\epsilon$. 
It is evident that $ \beta \le (1-2\epsilon)$ 
keeps the total number of clusters small.}
\label{fig:cluster_number}
\end{figure*}

We numerically evaluate the approximate upper bound for the number of clusters
$ T_N \approx T_{N-1} + P_N^*$. See Figure \ref{fig:cluster_number} for an application
of this analysis.


\subsection*{Commands}

Here we assume that the read files are named {\tt reads\_1.fq} and {\tt reads\_2.fq}.
TreQ-CG is a collection of three programs - {\tt treq-cluster, treq-split}, and {\tt treq-map}. 
At first we compute clusters, then extract anchor reads to be mapped by a traditional
read mapper, and finally, use the mapping information of anchor reads to find alignment of the
remaining reads.
The commands to run these programs are given below. 
Appropriate changes were made for single end libraries and for libraries with pre-Illumina 1.8 quality scores.

\begin{itemize}

\item Clustering reads:\\
  {\tt treq-cluster [-q 0] -t 24 -s 0.95 cluster\_prefix reads\_1.fq reads\_2.fq}
  
\item Extracting anchor reads:\\
  {\tt treq-split cluster\_prefix anchor\_prefix reads\_1.fq reads\_2.fq}
  
\item Mapping anchor reads:\\
\begin{itemize}
 \item {\bf BWA}\\
	{\tt bwa aln -t 24 index anchor\_prefix\_1.fq $>$ tmp\_1.sai }\\
	{\tt bwa aln -t 24 index anchor\_prefix\_2.fq $>$ tmp\_2.sai }\\
	{\tt bwa sampe -a 1000 index tmp\_1.sai tmp\_2.sai anchor\_prefix\_1.fq anchor\_prefix\_2.fq $>$ anchor\_reads.sam }
 \item {\bf Bowtie2}\\
	{\tt bowtie2 [--phred64] -x index -p 24 -S anchor\_reads.sam -1 anchor\_prefix\_1.fq -2 anchor\_prefix\_2.fq }
 \item {\bf Novoalign}\\
	{\tt novoalign -F [ILM1.8 | ILMFQ] -e 1 -r Random -o SAM -d index -f anchor\_prefix\_1.fq anchor\_prefix\_2.fq $>$ anchor\_reads.sam }
 \item {\bf Stampy}\\
        {\tt stampy [--solexa] -g index -h index -o anchor\_reads.sam -t 24 -M anchor\_prefix\_1.fq anchor\_prefix\_2.fq }
\end{itemize}

\item Clustered mapping:\\
 {\tt treq-map -t 24 genome.fa cluster\_prefix anchor\_reads.sam all\_reads.sam reads\_1.fq reads\_2.fq }
\end{itemize}

\begin{figure*}
  \centering
  \includegraphics[scale=0.6]{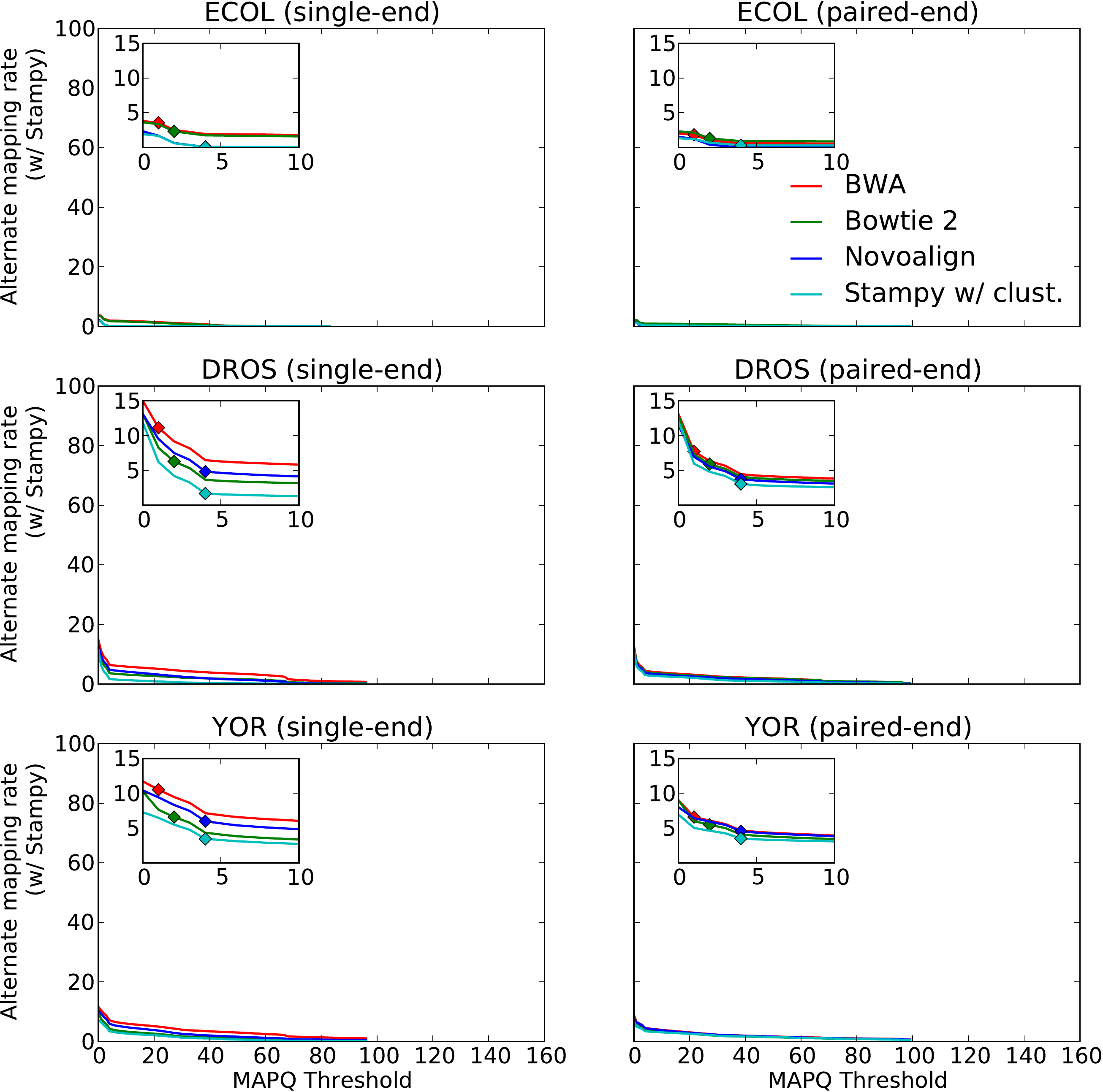} 
  \caption{ \label{fig:rm-vs-rm}
  {\bf Alternate mapping rate between readmappers compared to Stampy.} Alternate mapping rate is shown as a function of MAPQ threshold.}
\end{figure*}

\end{document}